# Effect of Low Temperature Baking in Nitrogen on the Performance of a Niobium Superconducting Radio Frequency Cavity

Pashupati Dhakal[1], Santosh Chetri[2], Shreyas Balachandran[2], Peter J. Lee[2] and Gianluigi Ciovati[1]

[1]Thomas Jefferson National Accelerator Facility, Newport News Virginia 23606, USA
[2]Applied Superconductivity Center, National High Magnetic Field Laboratory, Tallahassee, Florida 32310, USA

**Abstract**

We report the rf performance of a single-cell superconducting radiofrequency cavity after low temperature baking in a nitrogen environment. A significant increase in quality factor has been observed when the cavity was heat treated in the temperature range of 120-160 °C with a nitrogen partial pressure of ~25 mTorr. This increase in quality factor as well as the $Q$-rise phenomenon ("anti-$Q$-slope") is similar to those previously obtained with high temperature nitrogen doping as well as titanium doping. In this study, a cavity $N_2$-treated at 120 °C and at140 °C, showed no degradation in accelerating gradient, however the accelerating gradient was degraded by ~25% with a 160 °C $N_2$ treatment. Sample coupons treated in the same conditions as the cavity were analyzed by scanning electron microscope, x-ray photoelectron spectroscopy and secondary ion mass spectroscopy revealed a complex surface composition of $Nb_2O_5$, $NbO$ and $NbN_{(1-x)}O_x$ within the rf penetration depth. Furthermore, magnetization measurements showed no significant change on bulk superconducting properties.

## 1. INTRODUCTION

Recent advances in the processing of bulk superconducting radio frequency (SRF) niobium cavities via interior surface impurity diffusion have resulted in significant improvements in their quality factor ($Q_0$). The motivation for the development of these processes is to reduce the cryogenic operating cost of current and future accelerators while providing reliable operation. The potential for higher $Q_0$ in SRF cavities was first realized by titanium doping [1,2,3] during high temperature annealing of SRF cavity at 1400 °C and later by nitrogen doping at 800 °C [4], followed by electropolishing (EP). The material diffusion process not only resulted in an increase in quality factor at low field levels, but also an increase in quality factor with increasing accelerating gradient, contrary to the previously observed $Q$-slope. A possible explanation for the high quality factor is the trapping of hydrogen at interstitial sites due to the diffused atoms [5,6] and reduction on the formation of lossy hydrides during cavity cooldown [7]. A possible explanation for the $Q$-rise ("anti-$Q$-slope") phenomenon is related to the broadening of the peaks in the electronic density of states at the gap edges by the rf current [8,9]. However, despite the increase in $Q_0$, the quench field of the cavities doped by titanium or nitrogen is often limited to much lower values than achieved by standard treatments [3]. Even though no definitive cause is known, it is likely due to early vortex penetration due to the reduction of the lower critical field by material doping.

Results on SRF cavities which were heat treated in the temperature range of 600–1600 °C without subsequent chemical etching, showed improvements in the quality factor compared to standard preparation methods by minimizing both the residual and Bardeen-Cooper-Schrieffer (BCS) surface resistance [10,11]. Most recently, efforts have been made to preserve high accelerating gradients while also increasing the quality factor of SRF cavities [12,13]. In these new nitrogen "infusion" cavity processing recipes, cavities were heat treated at 800 °C for 3 hours and the furnace temperature is reduced to 120-200 °C and nitrogen is introduced into the furnace at a partial pressure of ~ 25 mTorr for ~48 h. This process has shown an improvement in $Q_0$ over the baseline measurements, without the need for post-furnace chemical etching. Even though diffusion of the nitrogen into the bulk of the SRF cavity is limited in depth at these low temperatures (120-200 °C), the introduction of nitrogen is sufficient to modify the cavity surface within the rf penetration depth as seen from rf results, which are similar to those previously reported for high-temperature nitrogen doped cavities. Furthermore, while post-doping electropolishing is required to remove coarse nitrides from the surfaces of high-temperature nitrogen doped cavities, no further processing is required for the low-temperature "infusion" recipe showing a clear benefit in reducing processing steps as well as keeping higher gradient with high $Q_0$ values. In this manuscript, we present the results from several rf tests on a single cell cavity treated in low temperature nitrogen environment as well as analyses of sample coupons treated under similar conditions.

## 2. CAVITY SURFACE PREPERATION

One single cell cavity of low-loss shape (frequency = 1.5 GHz, geometric factor = 277.2 Ω, $E_p/E_{acc}$ = 1.99 and $B_p/E_{acc}$ = 4.18 mT/(MV/m)) was fabricated from high purity (RRR>300) fine-grain Nb. The cavity fabrication followed the standard procedure of deep-drawing discs then machining of the iris and equator and finally electron beam welding the assembly. After the fabrication of the cavity, the interior was polished by standard electropolishing to remove ~150 μm from the inner surface of the cavity. The cavity was then subjected to high temperature degassing at 800 °C for 2 hours in a resistive vacuum furnace followed by the removal of 20 μm from the surface by electropolishing. Standard procedures were followed to clean the cavity surface in preparation for an rf test: degreasing in ultra-pure water with a detergent and ultrasonic agitation, high pressure rinsing with ultra-pure water, drying in the ISO4/5 cleanroom, assembly of flanges with rf feedthroughs and pump out ports and evacuation. The cavity was inserted in a vertical cryostat and cooled to 4.5 K with liquid helium using the standard Jefferson Lab cooldown procedure in a residual magnetic field of < 2 mG. This procedure results in a temperature difference between the two irises ΔT> 4 K when the equator temperature crosses the superconducting transition temperature (~9.2 K), which provides good flux expulsion conditions [14]. Before the heat treatment, the cavity was high pressure rinsed (HPR) and then dried in an ISO 4/5 cleanroom. While in the cleanroom, special caps made from niobium foils similar to those used in Ref. [12] were placed to cover the cavity flange openings. The cavity was then transported to the furnace inside a clean, sealed plastic bag. The vacuum heat treatment procedure started with the 800 °C/3 hrs degassing step followed by lowering the temperature to the (120-160 °C) range. The furnace is continuously pumped until the furnace reaches ~ 290 °C, at which point a nitrogen partial pressure of ~25 mTorr was introduced and this pressure is then maintained without active pumping of the furnace enclosure. Once the temperature has fallen to the desired value (120-160 °C), which is within ~2 hours, the temperature is held for ~46 hours. Figure 1 shows the typical temperature and partial pressure of the gas measured during the heat treatment process. After the heat treatment, the standard cavity cleaning procedures were applied before the rf test. After each heat treatment, the cavity's inner surface was reset (rf tests 4 and 6) with a standard ~10 μm EP. The sequential steps of cavity processing are as follows:

- rf test 1

- in situ 120 °C low temperature baking for 46 hours
- rf test 2
- HT at 800 °C/3 hrs followed by 120 °C/48 hrs in ~ 25 mTorr nitrogen
- rf test 3
- ~ 10 μm standard electropolishing
- rf test 4
- HT at 800 °C/3 hrs followed by 160 °C/48 hrs in ~ 25 mTorr nitrogen
- rf test 5
- ~ 10 μm standard electropolishing
- rf test 6
- HT at 800 °C/3 hrs followed by 140 °C/48 hrs in ~ 25 mTorr nitrogen
- rf test 7

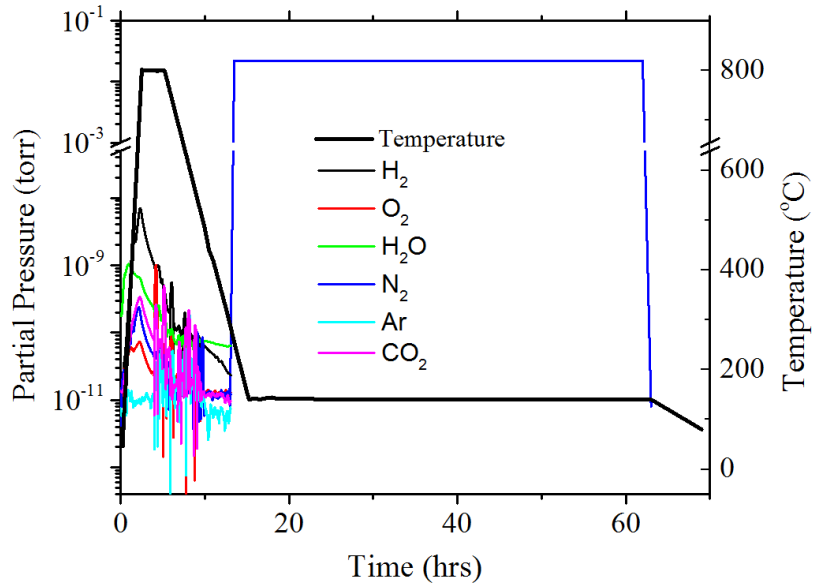

Figure 1. Typical temperature and partial pressure of the gas measured during the heat treatment process. All partial pressures during the heat treatment were below $10^{-8}$ torr. The residual gas analyzer was turned off before $N_2$ injection.

## 3. CAVITY RF RESULTS

The cavity immersed in the helium bath is excited using a phase-locked loop to measure the quality factor (surface resistance) as a function of temperature while the helium bath temperature was lowered by pumping down to ~1.5 K. The average rf surface resistance $R_s(T)$ was measured at low rf field ($B_p \sim 10$ mT) and fitted with the following equation

$$R_s(T) = R_{BCS}\left(T, \frac{\Delta}{K_B T_c}, l\right) + R_{res} \qquad (1)$$

where $R_{res}$ is the temperature independent residual resistance and $R_{BCS}$ is BCS surface resistance. Here, $\Delta$ is the superconducting energy gap at 0 K, $K_B$ is Boltzmann's constant, $T_c$ is critical temperature, and $l$ is the normal-electrons' mean free path. The BCS resistance was numerically calculated from the BCS theory of superconductivity, which is valid in the zero-field limit [15] with parameters $T_c = 9.2$ K, London penetration depth ($\lambda_L$) = 32 nm and coherence length ($\xi_0$) = 39 nm, considered to be material constants for superconducting niobium. Measurements of $Q_0(B_p)$ were done at 2.0 K up to the maximum gradient as limited by either breakdown or high field Q-slope.

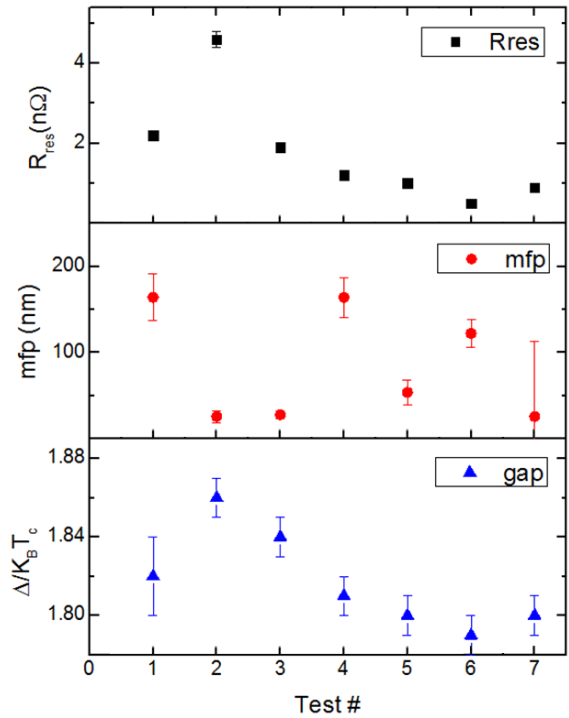

Figure 2. Material parameters obtained from the fit of $R_s(T)$ from 4.3 to ~1.5 K.

Figure 2 shows the $R_{res}$, $\Delta/K_B T_C$ and $l$ obtained from the fit of $R_s(T)$. The residual resistance is consistently $\leq 2$ n$\Omega$, except for test 2 when the cavity was subjected to low temperature baking (LTB) at 120 °C for 48 hours in vacuum. Although this increase in residual resistance and decrease in mean free path (mfp) has been typically observed in the past for LTB cavities [16], the explanation is still not resolved. The mfp after baseline EP treatments is significantly higher ($l\sim150$ nm) than after LTB and nitrogen infusion ($l < 50$ nm) is an indication, which is consistent with results from earlier measurements on cavities after LTB in vacuum [16].

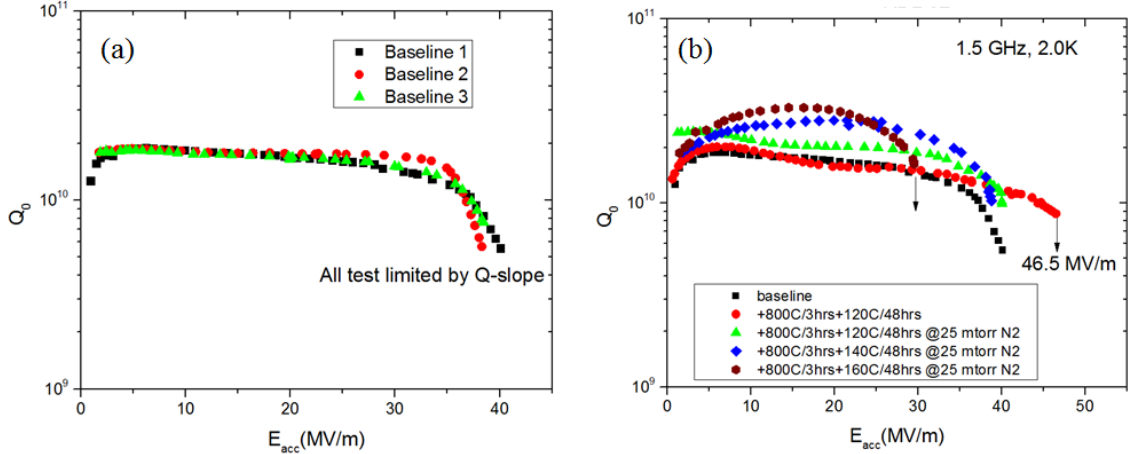

Figure 3. rf measurement taken at 2.0 K for (a) the baseline measurements and (b) after different surface treatments.

Figure 3(a) shows the $Q_0(E_{acc})$ curve for the three baseline tests (tests 1, 4 and 6) measured at 2.0 K. All these three baseline tests have similar $Q_0(E_{acc})$ dependence with small deviations above 30 MV/m, and all tests were limited by $Q$-slope without field emission. These results showed that ~10 μm surface removal by EP resets the surface such that a reproducible rf performance is achieved, regardless of the previous annealing. Figure 3 (b) shows the $Q_0(E_{acc})$ curves for the same cavity heat treated at different temperatures along with the baseline test 1. The cavity which received the LTB (test 2) at 120 °C for 48 hours on the vertical test stand after the baseline test 1 reached a maximum gradient of 46±2 MV/m, which corresponds to a peak magnetic field of ~ 196±8 mT, close to the critical field of Nb at 2.0 K. As observed in the past, the $Q$-slope at high field is also minimized by the LTB at 120 °C for 48 hours done under ultra-high vacuum (UHV) conditions. The cavity which was subjected to the additional 120 °C for 48 hours in a nitrogen partial pressure of ~25 mTorr showed no improvement in the accelerating gradient over the baseline electropolished cavity but a small increase in $Q_0$ (~20%) at all field levels. The improvement in $Q_0$ was not as high as recently reported on other cavities treated in a similar way [12], which may be due to not having a fresh electropolished surface prior to the furnace treatment. We plan to explore the effect of surface conditions on cavity performances prior to the heat treatments in the future.

The cavity heat treated at 800 °C for 3 hours followed by 160 °C for 48 hours in nitrogen partial pressure of ~25 mTorr showed an improvement of $Q_0$ by a factor of ~2, with an extended $Q$-rise (reaching a maximum at ~ 16 MV/m) as observed in earlier nitrogen doped cavities at higher temperatures (800 °C) [3,4]. However, the cavity quenched at ~30 MV/m with a $Q$-slope starting at ~20 MV/m. The cavity heat treated at 800 °C for 3 hours followed by a 140 °C for 48 hours nitrogen infusion showed the best performance, with increasing $Q_0$ to a broad peak and a maximum accelerating gradient of ~39 MV/m, similar to the baseline tests. The $Q$-rise phenomenon is probably similar to the high temperature nitrogen doped cavities [8,9], however, the high field $Q$-slope on these nitrogen infused cavities needs further investigation as the diffusion depth of Nitrogen into Nb is expected to be much less at low annealing temperatures.

## 4. COUPON PREPERATION AND CHARACTERIZATION

Samples (7.5 mm ×5 mm × 3.125 mm in dimensions) labeled F8-F12 were cut by wire electro-discharge machining (EDM) from a high purity fine grain niobium sheet. The samples were etched by BCP 1:1:1, removing ~70 μm, heat treated in a UHV furnace at 600 °C for 10 hours to degas hydrogen, and etched by BCP 1:1:2, removing ~30 μm. Afterwards, the samples were nano-polished at Wah Chang, USA, to obtain a surface with mirror quality smoothness. Sample labeled F8 is heat treated at 800°C/3h followed by 120°C/48h in UHV furnace. Sample F9 is heat treated at 800°C/3h, whereas samples F10, F11 and F12 were heat treated at 800°C/3h followed by 120°C/48h, 140°C/48h and 160°C/48h in nitrogen partial pressure of ~ 25 mTorr respectively. The treatment of these samples attempted to replicate the cavity treatments as much as possible, most importantly the samples were heat treated inside an Nb tube enclosed by Nb caps to simulate the environment with Nb cavity. The samples were analyzed by field emission scanning election microscopy (FESEM) to investigate the surface morphology and x-ray photoelectron spectroscopy (XPS) to identify the surface composition. Time of flight secondary ion mass spectroscopy (TOF-SIMS) was used for elemental information near the surface. Furthermore, cylindrical samples (M8-M12) with length ~ 3 mm and diameter ~1 mm were also heat treated along with the flat samples for dc magnetization measurement to determine the superconducting critical fields.

## I. Surface morphology

Surface imaging was performed with a Zeiss 1540EsB FESEM. The polycrystalline macroscopic surface of the coupon samples after different surface treatments are shown at low magnification in Figure 4 (a-d). FESEM back-scattered electron (BSE) images, which are sensitive to crystallographic orientation and composition, show a mottled contrast, suggesting residual surface strains, on all samples irrespective of the heat treatment condition. The overall grain sizes in all samples are similar and in the range 20-100 μm, which is typical of fine-grain Nb sheets. There is no change in the macro grain structure after different low temperature nitrogen infusion heat treatments between 120 °C-160 °C. Higher magnification BSE images of representative regions are shown in Figure 5(a-d). There is a clear difference in N-infused surfaces when compared with those annealed at the standard 800 °C/3h in UHV. There is very weak in-grain surface channeling contrast in 800 °C/3h, as shown in Fig.5 (a). All N-infused samples show surface structures as indicated by Figs. 5 (b-d). The density of the surface structures appears to increase with higher N-infusion temperature from 120 °C to 160 °C.

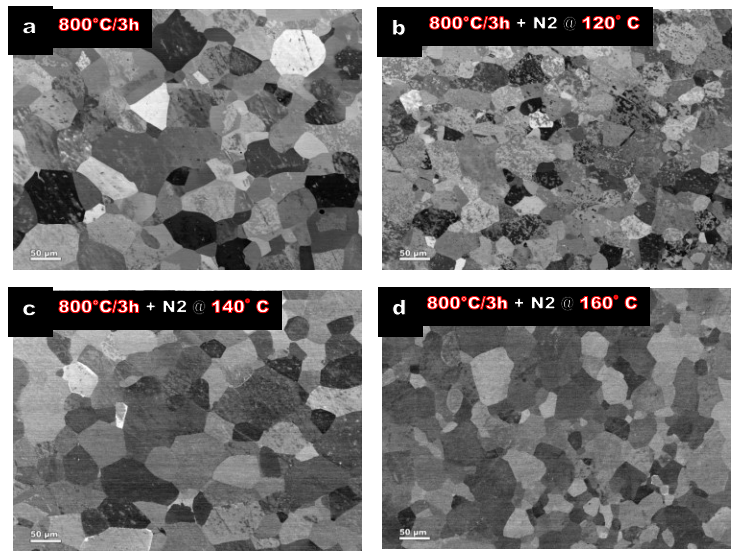

Figure 4. Representative low magnification BSE images coupon sample surfaces with grain boundaries (GB) after different heat treatments, (a) 800 °C/3h with no nitrogen infusion and (b)-(d) are low temperature nitrogen infused at 120, 140 and 160 °C respectively.

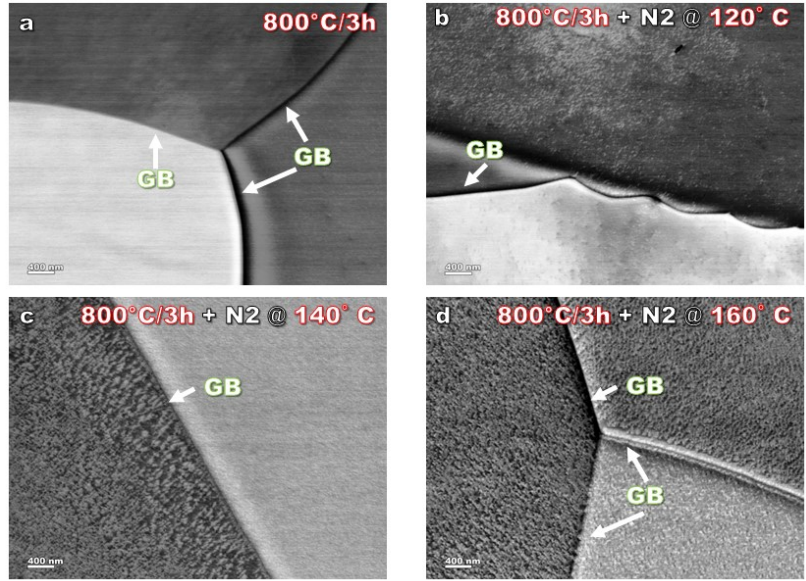

Figure 5. Representative high magnification BSE images coupon sample surfaces with grain boundaries (GB) after different heat treatments, (a) 800 °C/3h with no nitrogen infusion and (b)-(d) are low temperature nitrogen infused at 120, 140 and 160 °C respectively.

II. XPS Results

For the XPS study (Physical Electronics PHI 5000 Series) the photoelectrons were excited using an X-ray source that produces MgK$_\alpha$ radiation at 1253.6 eV. Before acquiring the data, a very light sputtering was done using 1 µA Argon ion at 3 keV just to remove any surface hydrocarbons. The data was acquired at different binding energy ranges for different elemental spectrum at a step size of 0.1 eV/step. The data was averaged among 10 cycles. The electron energy analyzer was operated in a constant energy mode with pass energy 71.55 eV. For angle resolved XPS the data was taken at different take-off angles: 15°, 30°, 45°, 60° and 75° (as the take-off angle of photoelectrons is increased the signal acquired is from an increased range of depth.

Figure 6 shows the Nb 3d peak for a sample that was heat treated at 800 °C /3h and 800 °C/3h with 140 °C N-infusion. The peak around 203.2 eV can be interpreted as $NbN_x$ or $NbN_{1-x}O_x$ peak. From the phase diagram of Nb-N, NbN is not thermodynamically favorable to occur below 400 °C, hence we interpret this peak mainly as a $NbN_{(1-x)}O_x$ peak. Also the Nb peak at 202 eV is absent in the nitrogen infused sample indicating that the $NbN_{1-x}O_x$ layer is sufficiently thick that the photoelectrons from the underlying Nb do not reach the surface. Furthermore, the Nb-3d peak intensities increase with take-off angle due to increased signal to noise ratio from the sub-surface. At the lowest take-off angle only the $Nb_2O_5$ peak is apparent. The $NbN_{1-x}O_x$ peak shows up only around 45 degrees take-off angle. This suggests that the surface of the low temperature nitrogen infused samples have a protective oxide layer, below which the oxy-nitrides layer is present with a thickness greater than 10 nm.

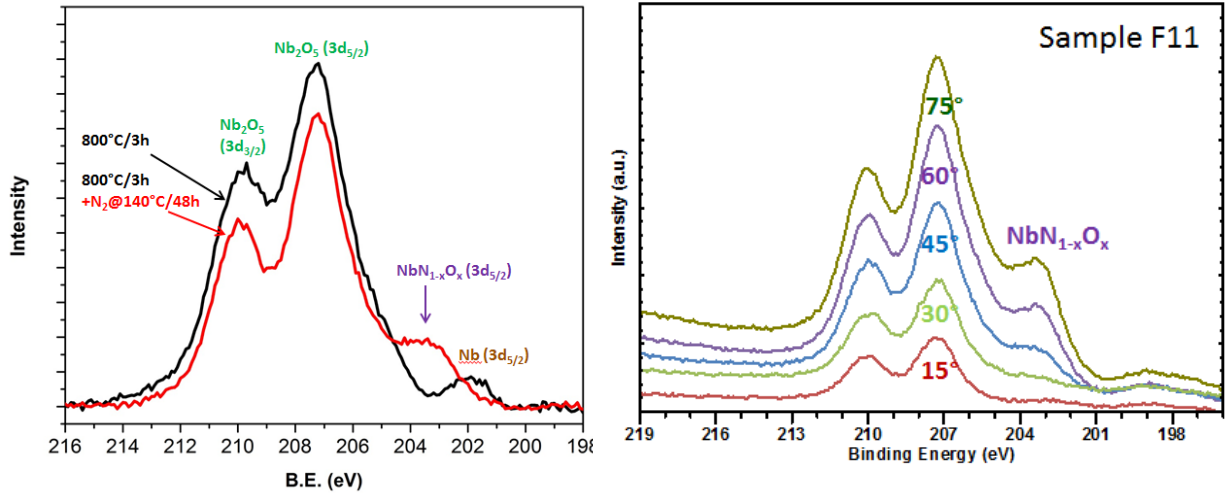

Figure 6. (a) Background subtracted XPS spectra on sample F9 (800 °C/3hrs) and F11 (800 °C/3hrs+140 °C/48hrs @ 25 mTorr $N_2$) at 45 deg take-off angle. (b) Angle resolved XPS on sample F11.

### III. TOF-SIMS Results

TOF-SIMS analyses [17] were conducted using a TOF SIMS V (ION TOF, Inc. Chestnut Ridge, NY) instrument equipped with a $Bi_n^{m+}$ (n = 1 - 5, m = 1, 2) liquid metal ion gun, $Cs^+$ sputtering gun and electron flood gun for charge compensation. Both the Bi and Cs ion columns are oriented at 45° with respect to the sample surface normal. The instrument vacuum system consists of a load lock for rapid sample loading and an analysis chamber, separated by the gate valve. The analysis chamber pressure is maintained below $5.0 \times 10^{-9}$ mbar to avoid contamination of the surfaces to be analyzed. For the depth profiles acquired in this study, 3 keV low energy $Cs^+$ with 20 nA current was used to create a 120 µm by 120 µm area, and the middle 50 µm by 50 µm area was analyzed using about 0.3 pA $Bi_3^+$ primary ion beam. The negative secondary ion mass spectra were calibrated using $H^-$, $O^-$, $Nb^-$, and $NbO^-$. The positive secondary ion mass spectra were calibrated using $H^+$, $Nb^+$, $CsNb^+$ and $Cs_2Nb^+$. The concentrations were calculated using standard C, N and O implant into standard niobium. Figure 7 shows the concentration of C⁻, O⁻ and NbN⁻ on samples F9-12. Since N does not have a significant negative secondary ion yield in SIMS, NbN⁻ ions were used to monitor the N-signal. As expected, sample F9 showed the lowest concentrations of O, C and N within rf penetration depth and elemental concentration increases with the temperature when nitrogen is injected during the furnace treatment. The concentration of oxygen increased with increasing baking temperature, most likely due to absorption of the residual oxygen from the furnace into the "unpassivated" niobium surface during the baking process. Higher oxygen concentration was found in the unpassivated niobium sample compared to the sample with surface passivation by nitrogen [10]. The higher concentration of NbN- observed in samples F11 and F12 indicates the diffusion of nitrogen within ~50 nm from the surface, which is consistent with the diffusion coefficient for N into Nb at 160 °C being $6.02 \times 10^{-25}$ m$^2$/s [18], resulting in a diffusion depth of 46 nm after 48 h. The NbN- concentration at a depth of ~10-50 nm from the surface is ~1-10 at.%

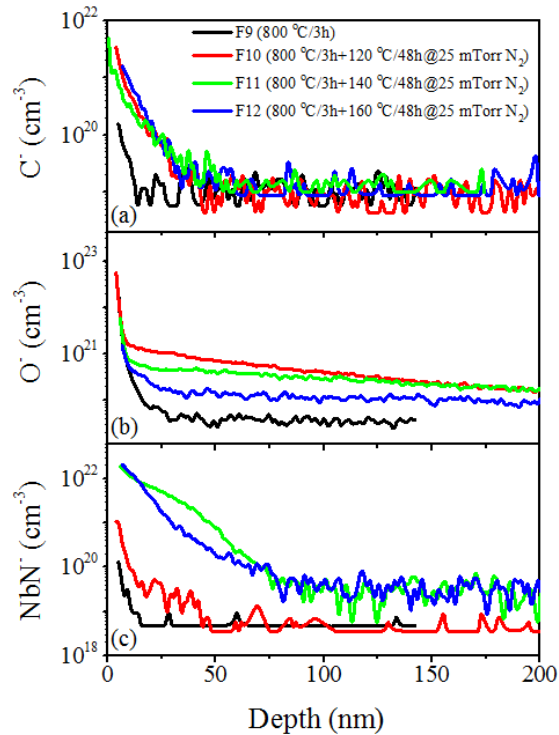

Figure 7. SIMS depth profile of (a) carbon, (b) oxygen and (c) nitrogen measured in samples F9-F12.

Figure 8 showed the ratio of counts/s for H⁻ and Nb⁻ for samples F9-F12. Surprisingly, the hydrogen concentration increased with increasing baking temperature. This increase may be due to either the reabsorption of hydrogen in the sample during the cooldown of the cavity from 800 °C or the trapping of the hydrogen near surface due to the introduction of nitrogen. Additional SEM measurements on cross-sectional areas of the samples are planned to further investigate the presence of hydrides close to the surface.

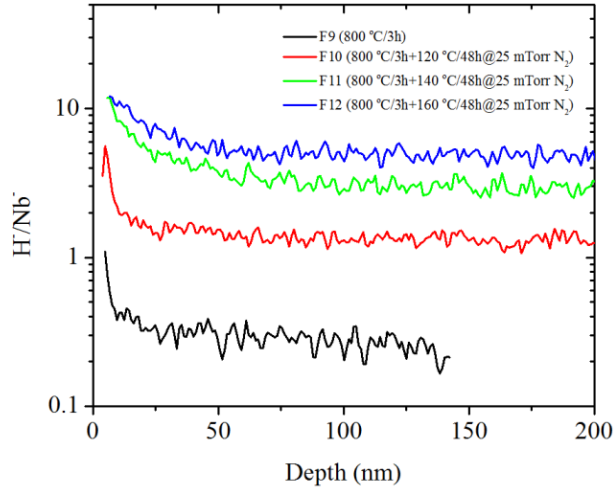

Figure 8. H⁻/Nb⁻ as a function of depth measured by ToF-SIMS in Nb samples F9-F12.

## IV. First flux penetration by DC magnetization

The DC magnetization measurements were acquired with 5 T Quantum Design SQUID (Superconducting Quantum Interference Device) system. Cylindrical samples were first zero field cooled to 4.2 K and then the dc magnetization measurements were made by applying a DC field parallel to the length direction of the samples. Figure 9 shows the isothermal DC magnetic hysteresis of samples M8-M12 measured at 4.2 K. No significant change in bulk pinning and field of first flux penetration ($H_{ffp}$) was observed after low temperature nitrogen infusion.

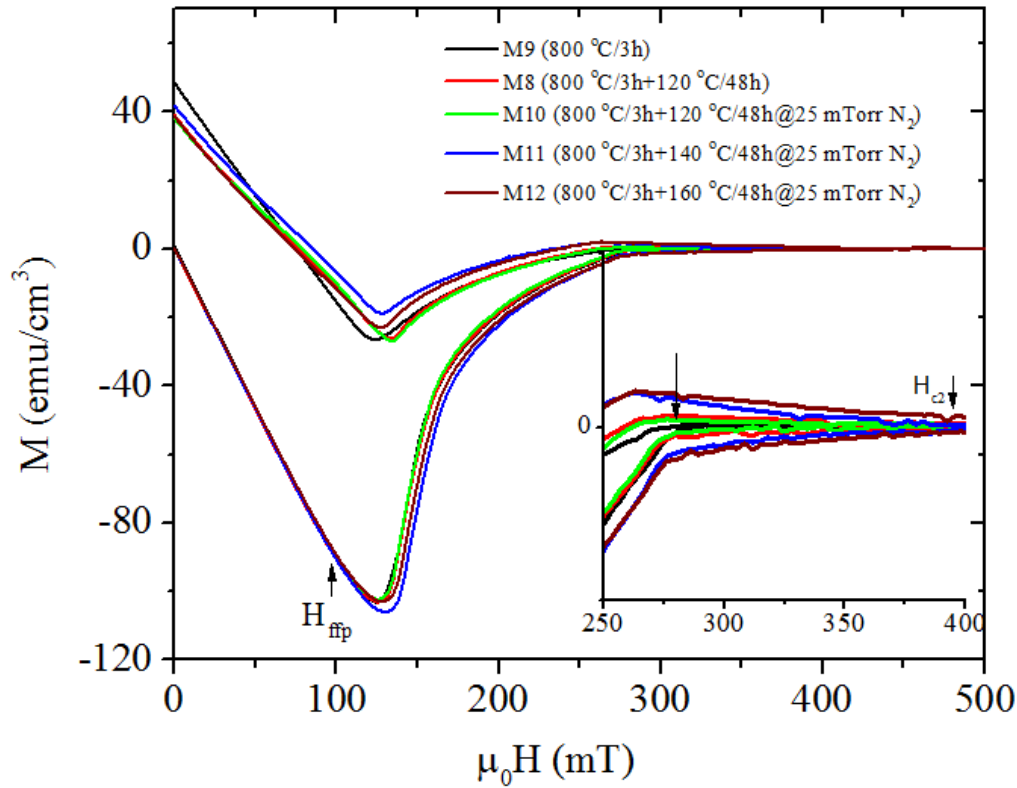

Figure 9. Isothermal DC magnetic hysteresis of samples M8-M12 measured at 4.2 K. The inset shows the *M(H)* close to $H_{c2}$ showing an enhanced $H_{c2}$ for samples M11 and M12.

## 5. DISCUSSION

The rf measurements on the treated SRF cavity and the surface analyses of the sample coupons showed that the heat treatment on SRF cavities at low temperature (120-160 °C) significantly alters the rf surface and hence the rf performance. Low temperature (100-150 °C) baking under ultra-high vacuum has been the standard practice for the final preparation of SRF cavities in order to recover from the high field *Q*-slope. The improvement in high field *Q*-slope with increase in accelerating gradient was explained as the result of oxygen diffusion in the bulk of the cavity [19] or due to the strong suppression of hydride precipitation caused by the change in concentration of vacancy-hydrogen complexes [20]. A comprehensive model capable of explaining all of the experimental results related to the high field *Q*-slope and baking effect is still lacking. Very limited data regarding the high-field behavior of cavities treated by the high-temperature nitrogen doping followed by EP, mainly because of the reduced quench field, typically < 100 mT, above which the high-field Q-slope typically occurs in BCP or EP treated cavities.

An important outstanding issue to increase the quality factor at or below 2 K is the increase in residual surface resistance, which is typically observed after the standard low temperature baking in UHV. Contributions to the residual resistance could come from the presence of "defects" such as damaged layers, metallic sub-oxides, sub-gap states, dielectric losses, trapped magnetic field.

XPS measurements indicate that the outermost dielectric $Nb_2O_5$ layer decomposes into metallic suboxides such as $NbO_x$ (x = 0.5 - 2) while baking at 120 °C in UHV [21, 22, 23]. However, subsequent exposure of the Nb surface to air re-oxidizes the surface increasing the thickness of $Nb_2O_5$ layer at the expenses of the sub-oxides [24]. RF measurements on cavities for which the oxide layer was stripped by rinsing with HF and re-grown after exposure to air and water, following the low-temperature UHV baking show that the residual resistance is reduced back to values similar to those prior to baking.

For the study reported in this article, the $Nb_2O_5$ is dissolved by the annealing at 800 °C and perhaps only few monolayers of $NbO_x$ might be on the cavity surface once the temperature dropped to <300 °C, at which point $N_2$ is injected into the furnace. A niobium oxy-nitrite layer might be formed on the surface during baking at 120-160 °C in nitrogen atmosphere and subsequent exposure of the surface to air and water promotes the growth of the outermost $Nb_2O_5$ layer. The XPS results indicate that an $NbN_{1-x}O_x$ layer still present between the $Nb_2O_5$ layer and the bulk Nb. The electronic properties of such layer and their influence on the electronic density of states of the adjacent superconducting Nb might explain the difference in the rf performance of "nitrogen infused" cavities compared to those which were subjected to the standard UHV baking. The role of hydrogen and the presence of hydrides in "nitrogen infused" cavities require further investigation.

A theoretical model in which the surface resistance of a superconductor coated with a thin normal metal was recently presented and showed that the $R_s(B_p)$ behavior observed in SRF cavities following different surface preparations can be explained with changes in the thickness of the normal layer and of the interface boundary resistance [25]. A recent theoretical model extends the zero-field BCS surface resistance to high rf fields [26]. Such model calculates $R_s(H)$ from the nonlinear quasiparticle conductivity $\sigma_1(H)$, which requires knowledge of the quasiparticles' distribution function. The calculation was done for two cases, one which assumes the equilibrium Fermi-Dirac distribution function and one for a non-equilibrium frozen density of quasiparticles. $R_s(H)$ is calculated numerically for these two cases and it depends on a single parameter, α, which is related to the heat transfer across the cavity wall, the Nb-He interface and between quasiparticles and phonons.

Figure 10 shows the measured $R_s(B_p)$ after N-infusion at 140 °C and 160 °C, normalized to the value at ~10 mT, along with the curves calculated with the model of Ref. [26]. Good agreement with the experimental data is obtained with α = 0.08 and the equilibrium distribution function for the N-infusion at 140 °C and with α = 1.125 and the non-equilibrium distribution function for the N-infusion at 160 °C. Considering a similar value of the thermal boundary conductance of ~1.7 kW/(m² K) at 2 K [27] for both N-infusion temperatures, the difference in the α-parameter between the two N-infusion temperatures implies that the quasiparticles-to-phonons energy transfer rate decreases from ~3.3 kW/(m² K) to ~0.08 kW/(m² K) as the N-infusion temperature increased by 20 °C. In both cases, electron overheating is the main bottleneck in the transfer of rf power into the He bath.

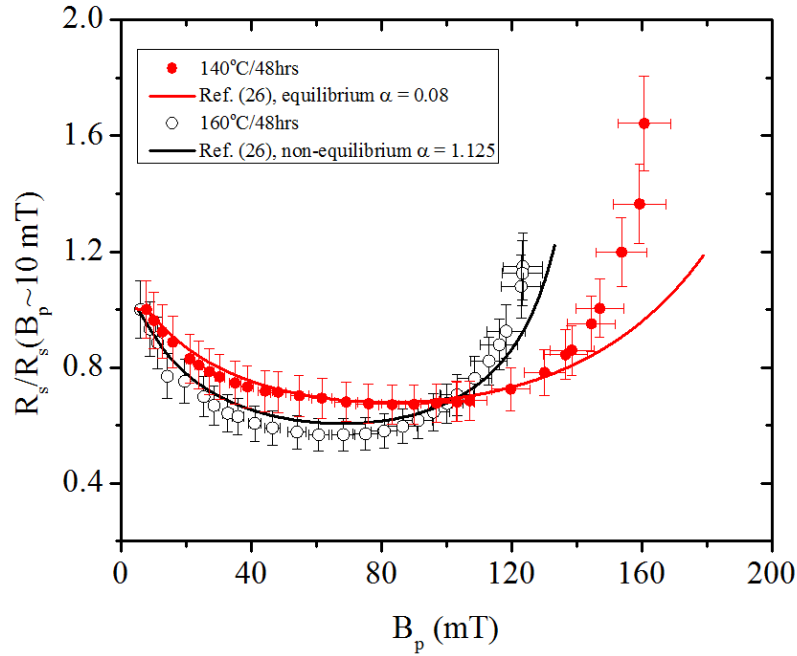

Figure 10. $R_s/R_s(B_p\sim 10mT)$ vs Bp calculated using the model in Ref.[26] along with the experimental data for cavity heat treated at 140 °C and 160 °C respectively.

The DC magnetization measurements showed no significant change on the bulk superconducting properties as a result of low temperature nitrogen infusion. The hysteresis loop observed in the magnetization data showed no change in the bulk magnetic flux pinning. A closer look near $H_{c2}$ (Fig. 6 inset) for samples M11 and M12 showed enhanced $H_{c2}$ values, probably due to the surface barrier effect [28,29] as a result of samples being treated at higher temperatures (140-160 °C) in nitrogen.

## 6. CONCLUSION

Improvement in the quality factor of an SRF Nb cavity was observed after annealing at 800 °C/3 h in vacuum followed by baking at 120-140 °C in low partial pressure of nitrogen inside a furnace ("N-infusion"), with a modest ~14% degradation of the maximum accelerating gradient. Larger reduction (~35%) of the quench field was observed when the baking temperature was 160 °C but with a pronounced increase of the quality factor with field, similar to that observed in cavities treated by high temperature nitrogen doping followed by electropolished.

SEM analysis of sample coupons showed the presence of surface features after N-infusion whereas XPS analysis showed the presence of a niobium oxy-nitrate layer at the surface. Impurities depth profiling by ToF-SIMS showed a diffusion profile for nitrogen in Nb down to ~50 nm and higher H concentration after N-infusion. Further studies are ongoing to better understand the role of impurities on the cavity performance.

The field dependence of the surface resistance after N-infusion can be described by a recent theoretical model that extends the calculation of the BCS surface resistance to high rf fields and indicates stronger electron overheating with increasing baking temperature. No significant changes of bulk critical fields or pinning properties seem to occur by N-infusion.

Overall, a quality factor as high as ~$2\times10^{10}$ at 1.5 GHz was achieved at a gradient of 35 MV/m by 800 °C annealing and N-infusion at 140 °C. Such performance would be of great interest for lowering the cryogenic heat load of high-energy accelerators such as the proposed Linear Collider [30].

## 7. ACKNOWLEDGEMENTS

We would like to acknowledge Jefferson Lab technical staff for the cavity surface processing and cryogenic support and Elaine Zhou at Analytical Instrumentation Facility (AIF), North Carolina State University for SIMS measurements. We would like to acknowledge Prof. A. Gurevich from Old Dominion University for useful discussion and providing the code to calculate the field dependence of the surface resistance. The work done at Florida State University is supported by the U.S. Department of Energy, Office of Science, Office of High Energy Physics under Award Numbers # DE-SC 0009960 (FSU) and DE-FG02-09ER41638 (MSU) and the State of Florida. Additional support for the National High Magnetic Field Laboratory facilities is from the NSF: NSF-DMR-1157490. This manuscript has been authored by Jefferson Science Associates, LLC under U.S. DOE Contract No. DE-AC05-06OR23177.